\documentclass[review]{elsarticle}
\usepackage{graphicx}
\usepackage{natbib}
\usepackage{subfigure}
\usepackage[citecolor=blue,colorlinks=true,linkcolor=blue]{hyperref}
\usepackage{dcolumn}

\begin{document}

\title{Off-lattice Pattern Recognition Scheme for Kinetic Monte Carlo Simulations }

\author{Giridhar Nandipati}
\ead{giridhar.nandipati@ucf.edu	}
\author{Abdelkader Kara}
\ead{abdelkader.kara@ucf.edu}
\author{Syed Islamuddin Shah}
\ead{islamuddin@knights.ucf.edu}
\author{Talat S. Rahman}
\ead{talat.rahman@ucf.edu}
\address{Department of Physics, University of Central Florida,  Orlando, FL  32816}

\date{\today}
\bibliographystyle{elsarticle-num}
\begin{abstract}
We report the development of a pattern-recognition scheme for the off-lattice self-learning kinetic Monte Carlo (KMC) method, one that is simple and flexible enough that it can be applied to all types of surfaces. In this scheme, to uniquely identify the local environment and associated processes involving three-dimensional (3D) motion of an atom or atoms, space around a central atom  is divided into 3D rectangular boxes. 
The dimensions and the number of 3D boxes are determined by the accuracy with which a process needs to be identified and a process is described as central atom moving to a neighboring vacant box accompanied by the motion of any other atom or atoms in its surrounding boxes. As a test of this method we apply it to 3D Cu island  decay on the Cu(100) surface and to 2D diffusion of a Cu monomer and a dimer on the (111) surface, and results and computational efficiency to those available in the literature.
\end{abstract}
\maketitle
\section{Introduction}
Quantitative understanding of nucleation and growth of heteroepitaxial films is challenging from both fundamental and technological points of view, as these are important processes\cite{r1,r2,r3} for the fabrication of nanostructures ranging from quantum wires\cite{Notomi} to quantum dots.\cite{Floro} Heteroepitaxial structures with strained semi-conductor thin films have found wide application in electronic and optoelectronic devices.\cite{kordos, Pearsall}
In many cases, strain due to lattice mismatch can lead to the formation of three dimensional (3D) clusters, \cite{Mo, Eaglesham} whose shapes can depend on variety of factors. 

A common method for simulating heteroepitaxial growth is to apply molecular dynamics (MD) using many-body interaction potentials. In the MD method, \cite{MD1,MD2,MD3} real motion of atoms is modeled through real-space integration of Newton's equations of motion and the forces acting upon each atom in the system are determined by the interatomic potentials for all atoms in the system.
One of  the main advantages of MD simulations is the explicit inclusion of system vibrational and thermal dynamics as controlled by the chosen interatomic potential. Together these allow the system to evolve freely (as a micro-canonical system) for timescales (a few hundred nanoseconds to microseconds) whose upper limit is imposed by available computational resources. MD simulations are in general limited to timescales of microseconds (the longest) that are still orders of magnitude smaller than those of experiments. 
Since typical experiments of interest (epitaxial growth, for example) report important morphological changes on timescales of minutes or hours, MD simulations may not be able to capture critical rare events and thereby fail to provide comprehensive evolution of the system.

Another method for studying epitaxial growth is to use kinetic Monte Carlo (KMC)\cite{bkl,gilmer,voter1,maksym,kristen,Blue} simulations.  
It is an extremely efficient method for carrying out non-equilibrium simulations of dynamical processes when the relevant rates are known. 
As a result, the KMC method has been successfully used to carry out simulations of a wide variety of dynamical processes over experimentally relevant time and length scales. In KMC the thermal motion of the system is included only implicitly and in an approximate way. In KMC, rates of allowed processes, through which the system evolves, are provided as an input. If this input is accurate and complete, KMC simulations are in a good position to be compared to experiments. One challenge is to determine as accurately as possible the parameters associated with these processes and, as completely as possible, the full list of possible processes for the system. Another challenge with KMC method is that the systems investigated must be discretized and mapped onto a fixed lattice in order to define various diffusion mechanisms that must be considered at a given moment. Heteroepitaxial systems are thus especially hard to treat with the KMC method because of the increased tendency for the system to go off-lattice, owing to strain due to lattice mismatch.

To address the problem of completeness, KMC methods have been developed that will find all the possible processes that can happen in the system on-the-fly,\cite{neb1,henkelman,slkmc1} removing the constraint that all the relevant atomic-scale events have to be known {\it a priori}. One such method is the self-learning KMC (SLKMC) technique.\cite{slkmc1} In SLKMC, a pattern-recognition scheme allows efficient storage and subsequent retrieval of information from a database of diffusion processes, their paths and their activation energy barriers. It has been used for detailed study of Cu island diffusion\cite{slkmc1,slkmc2} and of coarsening of Ag islands at late-stages \cite{pslkmc} and early stages \cite{iselect} on Ag(111) surface. This method is based on the assumption that all atoms sit on high-symmetry sites commensurate with the substrate (on-lattice sites) and are also at the same height. But for small clusters, atoms can sit on off-lattice sites even in homoepitaxial systems\cite{offkmc} such behavior is even more frequent for heteroepitaxial systems, in which atoms for islands of all sizes may occupy `off-lattice' sites. To describe these systems, a pattern-recognition scheme is required in which atoms are allowed to be at any position on the surface. In an effort to over come this problem Kara {\it et al} \cite{offkmc} developed two-dimensional (2D) off-lattice pattern recognition scheme and used it to study heteroepitaxial island diffusion. 
But that pattern-recognition scheme mentioned above can capture only 2D diffusion processes, a new pattern recognition technique is needed to model 3D motion of an atom. In this article we present a new off-lattice pattern-recognition scheme that can recognize three-dimensional (3D) processes and is flexible enough to be applied to all types of lattices. Although our original goal of off-lattice KMC method was to use it to study heteroepitaxial systems, in this article we test it by studying 2D diffusion of Cu islands on Cu(111) surface and decay of 3D Cu islands on Cu(100) and compare the results with those of previous studies.

The organization of this paper is as follows. In Section \ref{slkmc-p} we  give a very brief description of SLKMC algorithm. In Section \ref{pattern} we discuss the need for a new pattern recognition scheme and describe in detail our implementation of the new 3D off-lattice pattern-recognition scheme. In Section \ref{results}, as a test of our scheme we present results for diffusion of a Cu monomer and  a dimer on Cu(111) surface and for 3D Cu island decay on Cu(100) surface, and compare these with those of previously reported.  In Section \ref{conclusions} we offer some concluding remarks.


\section{A Self-Learning kinetic Monte Carlo Algorithm}\label{slkmc-p}
In on-the-fly KMC simulations, instead of using a fixed set of diffusion processes each with its activation barriers, all the possible diffusion processes are determined at each KMC step. In contrast, in SLKMC usage of a pattern-recognition scheme gives the ability to determine whether the energetics of all the possible processes have been determined and stored in a database during the course of a simulation.  If all the possible processes are stored in the database, no further action is taken, and the KMC simulation continues its course as in standard KMC with a complete list of processes. But whenever a new configuration is found, all its possible diffusion processes and their respective activation barriers are determined by saddle-point searches. The new configuration, together with its associated diffusion processes (each with its activation barriers) is stored in a database. Every new simulation starts from an empty database and gets filled up with new configurations and associated processes as it proceeds, until all the possible processes that can happen in the system are found.
 
 Several methods can be used to do a saddle-point search. A simple one is called the ``drag method,''  in which an atom is dragged in the direction of the nearest vacant site in small steps. During the process all the atoms in the model system are allowed to relax  in all directions, except for the diffusing atom which is allowed to relax only in the directions normal to that along which it is ``dragged''. This prevents the atom from moving back to the initial state. The method is efficient and easy to implement.  In future we will  be introducing more accurate methods for carrying out saddle point searches such as repulsive bias potential\cite{offkmc} and nudged elastic-band (NEB) method.\cite{neb1,neb2}


\section{2D Pattern-RecognitionScheme} \label{pattern}
As mentioned earlier, in the KMC method we need to know all the processes possible. Each process involves  some particular motion of an atom (or atoms), the activation barrier for which depends on the local neighborhood. In order to uniquely identify the local neighborhood and the diffusion processes associated with, and to store and retrieve this information on the fly, a pattern-recognition scheme is necessary.  Storing and retrieving this information avoids redundancy, as the system ``learns'' from its ``past,'' that is, from its memory of  the processes and their energetics associated with previously encountered patterns (or configurations). In order to succeed, a pattern-recognition scheme must be accurate enough to uniquely distinguish different shapes in the system as they appear and to make appropriate decisions using information already stored.

One of the first pattern-recognition schemes \cite{slkmc1} developed was designed to model on-lattice  2D diffusion of atoms on a (111) surface (It can be easily extended to other types of surfaces). In this scheme, neighboring fcc sites around a central atom are grouped into rings and, depending on occupancy of fcc sites in these rings, a decimal integer is generated that uniquely identifies the neighborhood around the central atom. Though very simple and easy to implement, this scheme cannot accommodate systems in which an atom sits at an hcp site or a non-symmetric (off-lattice) position, as occurs under conditions of lattice mis-match. To allow for off-lattice sites, a new pattern-recognition scheme \cite{offkmc} was developed to handle 2D diffusion of clusters on (111) surfaces (It too can be easily extended to other types of surfaces). In this scheme, a set of relative positions of atoms with respect to a chosen `reference point' is used to identify a local neighborhood around a leading or central atom. For the fcc (111) surface, the reference point is either (1) a fcc  or hcp site closest to the `leading atom' (any atom in the system can be selected as the leading atom) in the cluster or (2) a fcc or hcp site closest to the center of mass of the cluster. Every atom in the cluster then has uniquely defined coordinates with respect to this reference site.  Instead of an integer configuration key, each atom's distances in x, y \& z directions with respect to the reference site are stored in the database along with each of its possible diffusion processes and their respective activation barriers. We note that these distances of atoms from the reference point are not stored in any particular order in the database.

Although the latter way of classifying the neighborhoods overcomes restriction of the method to on-lattice systems, one of its main disadvantages is that the identification of pattern involves matching non-integer distances of atoms in the simulation with those in the database. Tolerance for matching these real (non-integer) numbers becomes very important: if the tolerance is small, the database will be very large (with the result that configurations might be redundant);  if it is large, a valid configuration might not be identified as unique.  In the database each configuration records the  shape of an island in terms of the relative distances of each atom with respect to the chosen reference point (whether a  fcc or a hcp site). All the possible processes and their activation barriers are attached to the configuration. Computational effort for pattern recognition increases with cluster size, since it involves matching the relative distances of all the atoms in a cluster with those stored in the database.
 For an individual configuration with an unsorted list of distances, computational effort for matching distances of each individual atom in the cluster increases as square of the number of atoms in the cluster. For example, for an $N$ atom cluster, in order to match the coordinates of the first atom, the simulation does $O(N)$ searches, for the second atom $O(N-1)$ searches and so on. Hence computational effort in matching a pattern  increases approximately as $N^{2}/2$ with the cluster size.


\section{3D Off-lattice Pattern Recognition Scheme}\label{3D}

 \begin{figure}[h]
\center
\subfigure{ \includegraphics [width=3.5cm]{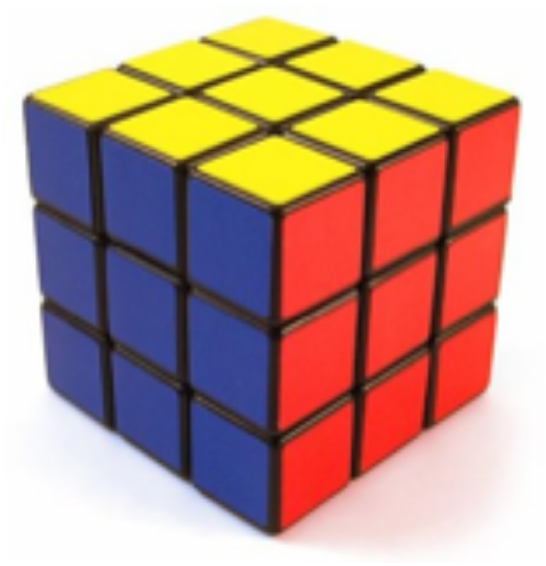} }
\caption{\label{3Dbox}{3D space is split up into 3D boxes with central atom at the center of the middle box.}}
\end{figure}

\begin{figure}
\center
\subfigure{ \includegraphics [width=7.2cm]{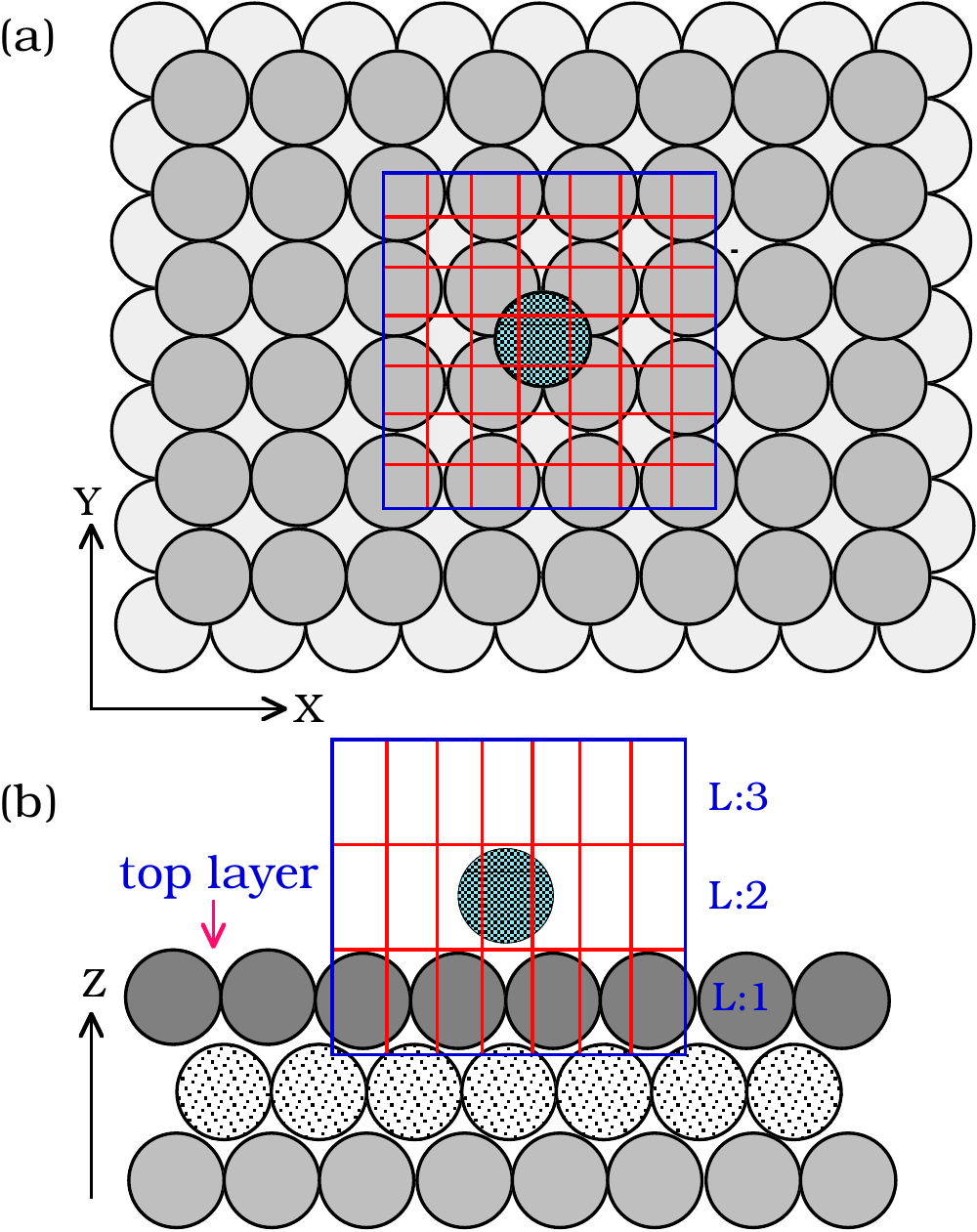} }
\caption{\label{boxfcc}{Monomer on fcc (100) surface (a) top view (b) side view. The bigger box (blue) is called the ``super box''; smaller ones are called just smaller boxes or occasionally boxes. L1, L2, L3 are layers}}
\end{figure}

To overcome the shortcomings of previously developed pattern-recognition schemes, a new scheme is needed which, apart from being applicable to off-lattice systems, should have following properties: (1) pattern matching should be based on integers that identify the neighborhood of a central atom; (2) it should be flexible enough to be applicable to all lattice types; and (3) it should be able to handle both 2D and 3D diffusion processes. In the new scheme developed here, 3D space around the central atom is divided into  rectangular boxes of appropriate size as shown in Fig.~\ref{3Dbox}, which looks like a Rubik's cube from the outside. The size of ``super box'' (the blue box in Fig.~\ref{boxfcc}) that encloses the smaller rectangular boxes (we call them simply ``boxes'') is dependent on the range of interaction in the system being studied. We note that  the neighborhood of every atom in the system is  identified by treating it as a ``central atom.'' The entire system of boxes is designed so that this central atom is always at the center of the middle box; thus the middle box is always occupied while the neighboring atoms are assigned to a given box if the center of mass lies within that box, allowing them to be anywhere within a box.  Based on the occupancy of the constituent  boxes, a unique binary number is obtained to identify the particular neighborhood around the central atom. 
A process occurs whenever that central atom moves into a neighboring box (anywhere  within the box). Individual processes are distinguished by which neighboring box that atom moves into and by particular accompanying movements of other atoms within the super box. 
If some other atom moves without the central atom's moving, that movement does not count as a processes for the central atom itself, but only for that other atom, when considered as central in its own right. 

\subsection{Implementation}
In this article we use the fcc($100$) surface to fully describe the implementation of our pattern recognition scheme. Figs.~\ref{boxfcc}(a) \&(b) show the top and side views of a 3D box system. The idea is very simple: a 3D box is constructed around the central atom symmetrically, whose dimensions depend on the range of the interaction for the system understudy. Consider the blue box as a 3D box around the central atom (blue-colored atom in  Fig.~\ref{boxfcc}), which is at the center of this 3D box, we call this box the ``super box.'' This super box is further divided into equal-sized smaller boxes. 
Division of the super box into smaller boxes is done in such a way that all boxes are distributed symmetrically around the central atom, assuring that it is always at the center of the middlemost box. 
These smaller boxes need not have same the dimensions in x, y, \& z directions as long as the dimensions are kept the same for all boxes. The size to be chosen depends on the precision with which a process needs to identified. Neighboring atoms are assigned to one of these smaller boxes on the basis of whether its center of mass falls within the box. Unlike the central atom, which is under the restriction of being always at the center of the middlemost box, neighboring atoms can be anywhere within the box so long as only one atom is allowed in each box. A binary number can then be generated based on the occupancy (1 for occupied and 0 for unoccupied), then converted to a decimal and stored in a database along with the information concerning all possible processes for each configuration and their respective activation barriers, as we will discuss in detail in the Section \ref{database}.


For the case of fcc ($100$) surface, the size of the super box is chosen so as to include at least $3$ layers of atoms (see fig.~\ref{boxfcc}(b)), a substrate layer $L1$ and adsorbate layers $L2$ \& $L3$. Depending on whether the system being studied is 2D or 3D, layer $L3$ will be either empty or occupied. We divide the super box into $7 \times 7 \times 3$ smaller boxes in the x, y \& z directions respectively, which corresponds to $49$ boxes in each of the layers in x-y plane, and a total of $147$ boxes in all $3$ layers. These numbers can be adjusted according to need. To uniquely identify the 3D neighborhood, a binary number is generated based on the occupancy of the boxes. For a $147$-box system, that will be a $147$-bit binary number, which is too large to handle. This binary number is thus split into $49$-bit numbers which represent the occupancy of boxes in each of the three layers. Thus, the 3D neighborhood around a central atom is uniquely represented by $3$ decimal integer numbers, which we call them ``layer numbers''. As mentioned earlier, the number of these small boxes is an adjustable parameter. As can be seen in Fig.~\ref{boxfcc}, the height of the small box has been set equal to the interlayer spacing, while the width and length are set to half the nearest-neighbor distance sufficiently small to accurately distinguish among $4$-fold, hollow, bridge and A-top sites. 

\begin{figure}
\center
\subfigure{ \includegraphics [width=7.5cm]{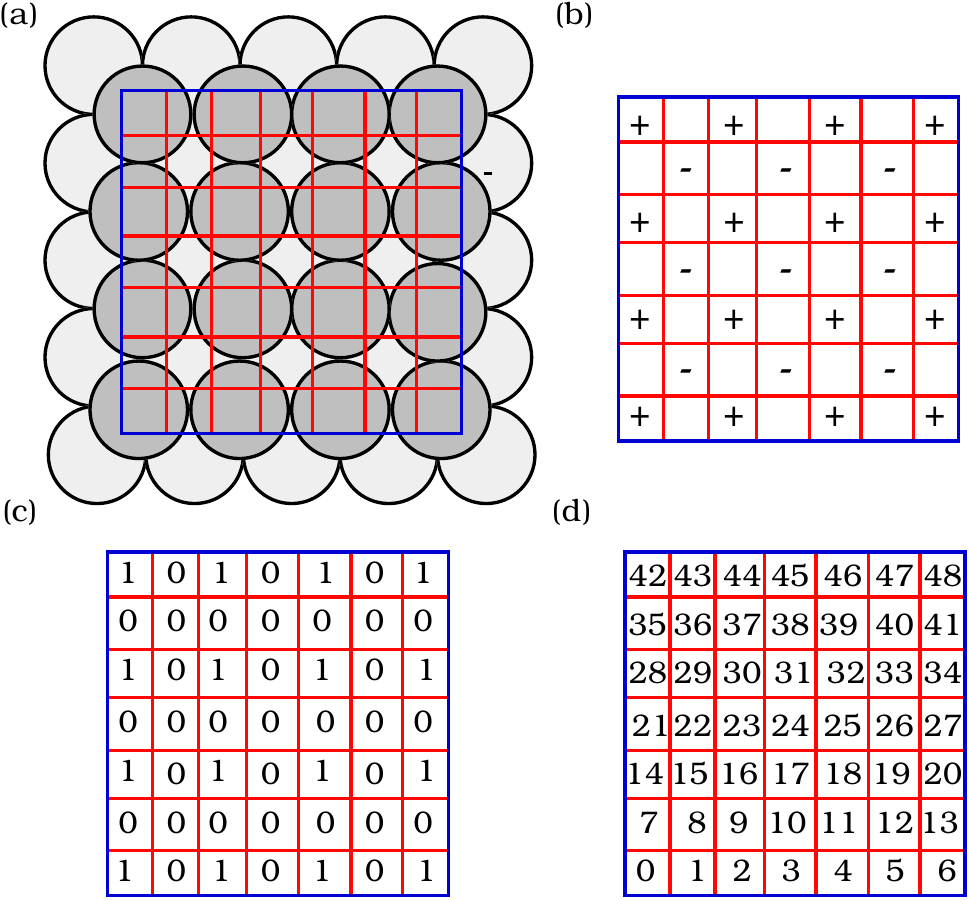} }
\caption{\label{boxfcc_bits}{Generation of a binary number for the top most substrate layer $L1$. (a) Top view of boxes in this substrate layer (b)  sign ``+''  represents the presence of a substrate atom, ``-'' represents a four-fold hollow site, and empty box represents a bridge sites (c) 0's and 1's for the substrate layer (d) Order in which  binary digits are read for each layer.}}
\end{figure}

\begin{figure}
\center
\subfigure{ \includegraphics [width=5.0cm]{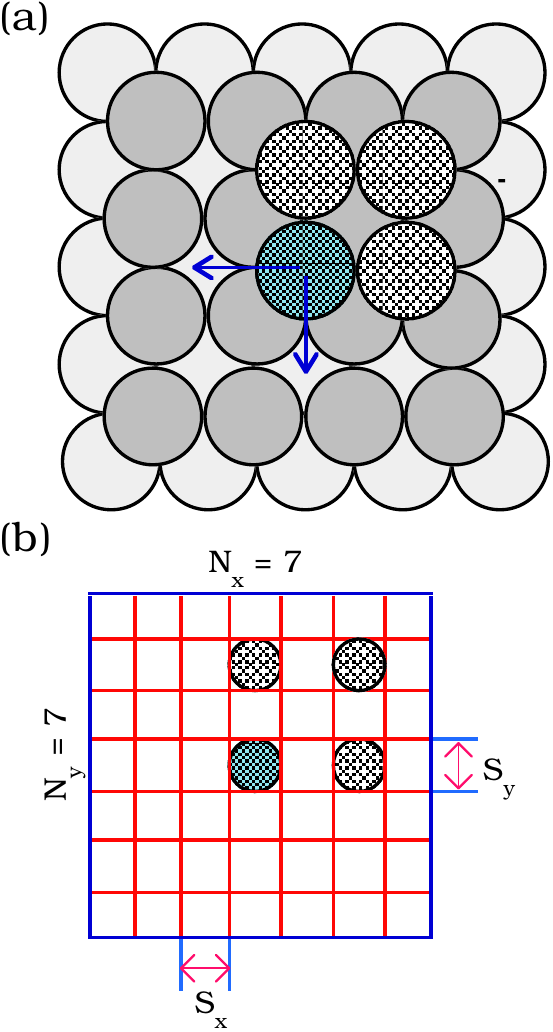}}
\caption{\label{dbconfig}{(a)Tetramer on a fcc(100) surface. Green colored circle repents the central atom. (b) location of atoms in the box system in layer $L2$}}
\end{figure}

To generate the binary number that uniquely identifies the 3D neighborhood around the central atom, boxes that are filled by the neighborhood atoms are identified and are assigned 1's while rest of the boxes are assigned 0's.  Fig.~\ref{boxfcc_bits} shows the generation of a layer number for the top most substrate layer $L1$. By comparing Figs.~\ref{boxfcc_bits}(a) \& (b) it can be seen that box with ``+''  represents box with the substrate atoms, ``-''  represents boxes with four-fold hollow sites and empty boxes represent bridge sites. All boxes, except the ones with ``+'' sign, are empty.  Accordingly Fig.~\ref{boxfcc_bits}(c) shows 1's and 0's based on the occupancy of the boxes for the substrate atoms. Digits of the binary number in each layer are read starting from box \# 0 to box \# 48 as shown in the Fig~\ref{boxfcc_bits}(d).  For the substrate layer $L1$ (dark-colored circles) shown in fig~\ref{boxfcc}(a), layer number is equal to $2^0+ 2^2 + 2^4 + 2^6 + 2^{14} + 2^{16} + 2^{18} + 2^{20} + 2^{28} + 2^{30} + 2^{32} + 2^{34} + 2^{42} + 2^{44} + 2^{46} + 2^{48} = 373856771850325$. For the case of an atom (green colored circle) in a tetramer in fig.~\ref{dbconfig}(a) in the layer $L2$, layer number is equal to $2^{24} + 2^{26} + 2^{38} + 2^{40}= 13744734208000$ (map Fig.~\ref{dbconfig}(b) onto Fig.~\ref{boxfcc_bits}(d)). For the empty third layer $L3$, it is zero.  These three numbers are then stored in the database in a pre-determined format which we will discuss in Section \ref{database}.

\begin{figure}
\center
\subfigure{ \includegraphics [width=4.0cm]{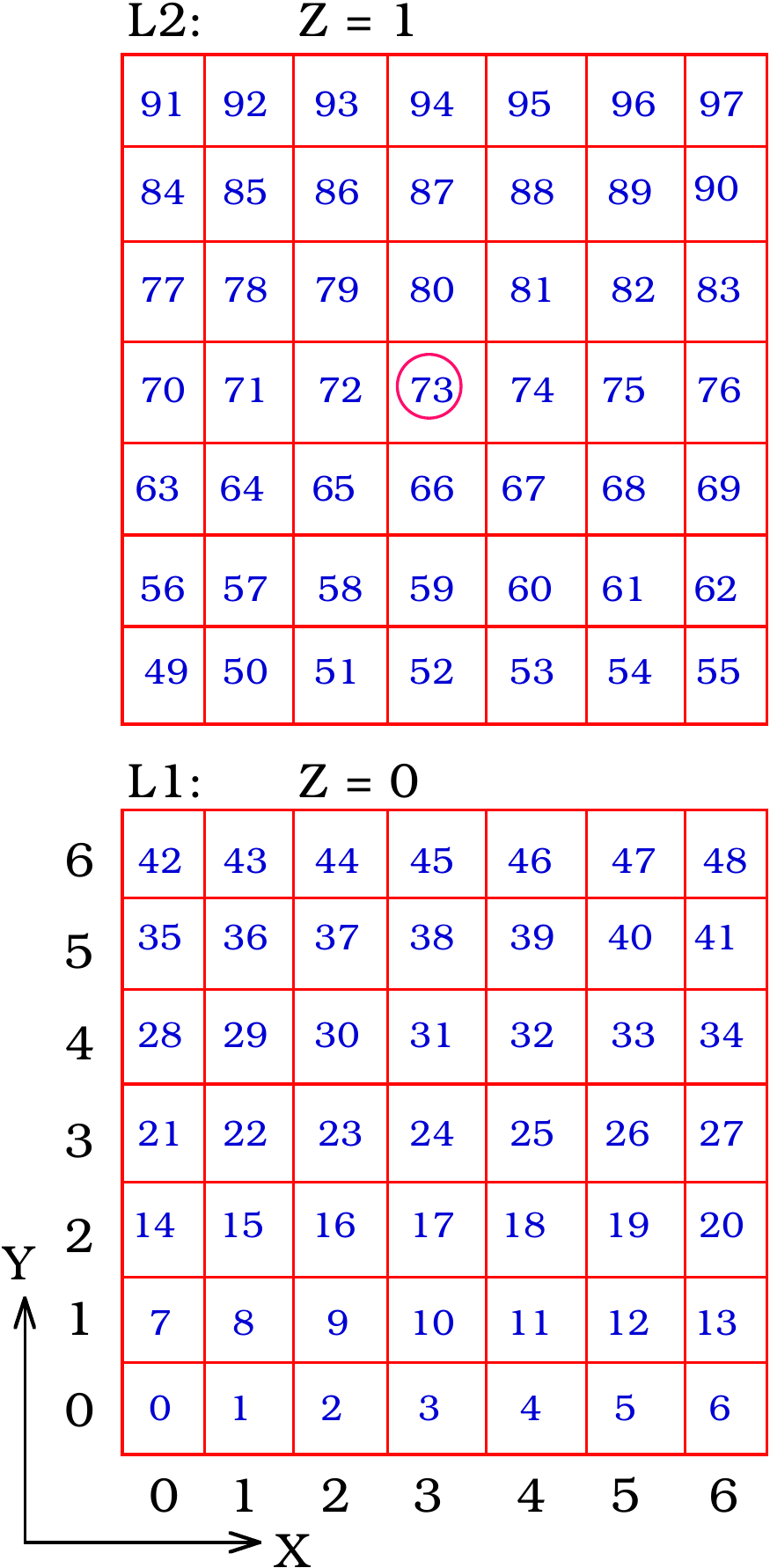}}
\caption{\label{layers}{Numbering of smaller boxes in the first two layers. Box with red circle is the location of central atom. Also shows integer coordinate system in which each box is a lattice point}}
\end{figure}

To implement this method all that is necessary is a way to identify the number of the box where neighboring atoms  are located. A method can be easily developed to extract these box numbers from their positions relative to the central atom. Fig.~\ref{layers} shows the sequential numbering of smaller boxes starting for the bottom two layer $L1$ \& $L2$ for the purpose of their identification. Numbering of boxes starts from the left corner box (Box \# 0 in Fig.~\ref{layers}) in the bottom most layer, which is $L1$ and continues sequentially to the right corner box (box \#146, not shown) in the third layer $L3$.  We note that this sequential numbering of boxes shown in Fig~\ref{layers} is independent of numbering in Fig.~\ref{boxfcc_bits}(d) showing the order in which binary digits are read for each layer. These box numbers can be obtained from integer coordinates, with Box \# $0$ being the origin as shown in Fig.~\ref{layers} (each box is a lattice point in this integer coordinate system).  If we know the integer coordinates of a box with respect to Box \# 0 then its box number is given as 
\begin{equation} 
{\rm Box~ Number} = i + j*N_{x} + k*(N_{x}*N_{y}) \label{boxnum}
\end{equation}
Accordingly, Box \# $0$ has integer coordinates of $(0,0,0)$, Box \# $73$, which holds the central atom has coordinates $(3,3,1)$. 
These integer coordinates can be obtained from the relative distances of neighboring atoms in the x, y \& z directions with respect to the central atom.
 If $x_{c}, y_{c}$ \& $z_{c}$ are the real coordinates of central atom and $x_{n}, y_{n}$ \& $z_{n}$ are the real coordinates of a neighboring atom, then integer coordinates are obtained by the use of following simple equations.
\begin{eqnarray}
i = \Big(\frac{x_r}{s_{x}}\Big)_I +\Big(\frac{N_{x}}{2}\Big)_I  ~~~~ x_{r} = x_{c}-x_{n} \\ 
j = \Big(\frac{y_r}{s_{y}}\Big)_I +\Big(\frac{N_{y}}{2}\Big)_I  ~~~~ y_{r} = y_{c}-y_{n}\\
k = \Big(\frac{z_r}{s_{z}}\Big)_I+\Big(\frac{N_{z}}{2}\Big)_I ~~~~ z_{r} = z_{c}-z_{n}
\end{eqnarray}
$\big(\big)_I$ represents {\it integer division} in which any fractional part or reminder is discarded. $x_r, y_r, z_r$ are real coordinates of a particular neighboring atom relative to the central atom, i.e., the difference in the $x, y ~\& ~z$ coordinates of central atom and those of neighboring atom.  While $s_{x}, s_{y}$ \& $s_{z}$ (see fig.~\ref{dbconfig}(b)) are dimensions of  a smaller box  and $N_{x}, N_{y}$ \& $N_{z}$ are the number of small boxes in $x, y ~\& ~z$ directions respectively; then $i$, $j$, and $k$ are integer coordinates of the small box in which that neighboring atom is located with Box \# $ 0$ being the origin.  Since our super box is divided into $7\times7 \times 3$ smaller boxes, $N_{x} = 7$, $N_{y} = 7$ \& $N_{z} = 3$ as shown in fig.~\ref{dbconfig}(b). And since the surface is Cu$(100)$, $S_{x} = 1.28$\AA, $S_{y} = 1.28$\AA  ~and $S_{z} = 2.08$\AA. we note that there is no particular relationship between the dimensions of the small box and lattice spacing except that the former have to be smaller than the latter and small enough to allows only one atom in each box (so that displacement of atoms moves them from one box to another). 

For the case of a tetramer shown in Fig.~\ref{dbconfig}(a) all four atoms are in the layer $L2$. Their integer coordinates if the blue-colored circle in Fig.~\ref{dbconfig} is the central atom are, $(3,3,1)$ for the central atom  and, $(5,3,1)$, $(3,5,1)$ and $(5,5,1)$ for the remaining three atoms. Accordingly the box number are obtained by using Eq.~\ref{boxnum}, which are $73$ (central atom), $75$, $87$ and $89$. Within the layer $L2$ these four atoms are in boxes $24$, $26$, $38$ and $40$, which are obtained by subtracting $49$, the number of boxes in the layer $L1$.
By following this method of checking the neighborhood within the range of interaction one can obtain the list of occupied boxes, thereby generating a decimal number that uniquely identifies the neighborhood.

In contrast to an on-lattice KMC simulation,  in an off-lattice KMC a neighborhood list used in order to identify the neighborhood of any atom efficiently.  To do this  we divide the system into equal-sized cells whose dimensions are larger than the range of interaction and a list of all atoms with each cell is compiled, similarly to what is done in molecular dynamics (MD) simulations.\cite{MD-book} To identify the neighborhood of an atom, all the atoms  within the cell where this atom is located are checked, instead of the entire system. In order to accommodate situations where an atom is near a cell boundary, the list also includes information about atoms from the adjacent cells that are within a distance of half the cell size.

\subsection{The Database}\label{database}

\begin{figure}
\center
\subfigure{ \includegraphics [width=7.5cm]{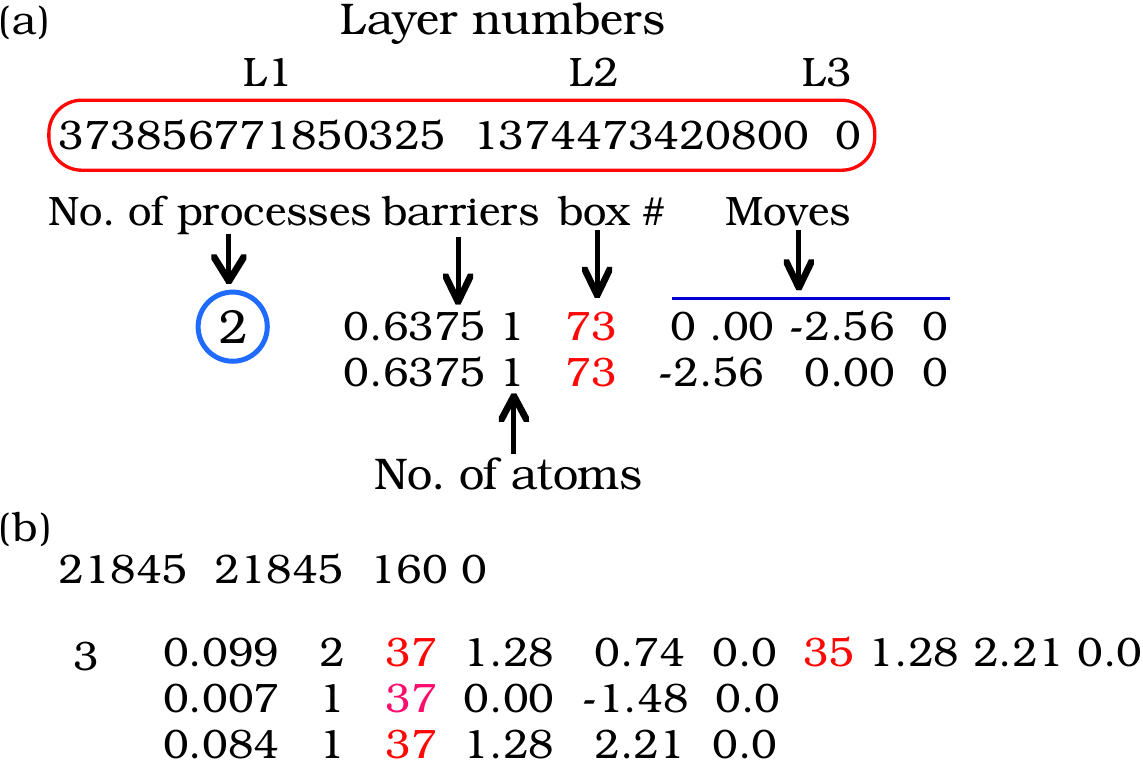}}
\caption{\label{dbformat}{Format of the database (a) Configuration for an atom (blue-colored circle) in a tetramer shown in Fig.~\ref{dbconfig} on a fcc(100) surface. (b) Configuration for an atom in a dimer on a fcc(111) surface.}}
\end{figure}

Three integer layers number are used to determine the configuration on a fcc(100) surface, similar to three ring numbers used in the original SLKMC on a fcc (111) surface.
Fig.~\ref{dbformat} shows the format through with each configuration and its  associated processes are stored in the database. The configuration shown in fig.~\ref{dbformat}(a) is for a central atom (blue-colored circle in fig.~\ref{dbconfig}(a)) in a tetramer on a fcc(100) surface, that shown in  fig.~\ref{dbformat}(b) is for an atom in a dimer with both atoms on fcc sites on a fcc (111) surface. In  fig.~\ref{dbformat}(a), the first three numbers (inside the red circle) are the layer number for each of the three layers. The next number (inside the blue circle) is the number of processes   possible for that configuration.  Individual processes are characterized by their activation barrier, the number of atoms involved in the process, the box \#'s (colored red) of those atoms and  displacements of these in x, y \& z directions. The format of the database for the fcc(111) surface is similar except that we use four layer numbers to identify the configuration, which comprises of two layers below the central atom and one layer above it.  We note that while the pattern-recognition acts upon integers instead of real numbers, the actual motion of the atoms are described in real numbers, which are displacements in x, y \& z coordinates from the current coordinates of the adatoms.

In order to minimize the size of the database, we exploit the symmetry of the type of the lattice under study. For the  case of fcc(100) surface we used (1) $90^{0}$ rotation, (2) $180^{0}$ rotation, (3) $270^{0}$ rotation, (4) mirror reflection, (5) mirror reflection followed by $90^{0}$ rotation, (6) mirror reflection followed by $180^{0}$ rotation, and (7) mirror reflection followed by $270^{0}$ rotation. If a given configuration is not found in the database, then symmetry operations are performed to find symmetric configurations and a new search is carried out to discover whether any of these exist in the database. This way of finding them on-the-fly saves memory at the expense of computational time.  Instead of performing them when required, symmetric configurations can be found ahead of time and stored in the memory during the simulation.  
Still, for small databases, where the memory is not an issue, storing symmetric configurations in the memory actually improves the speed of the simulation.  

Every time  an unknown configuration (representing a local neighborhood not previously met with) is encountered, symmetry operations are performed and the database is searched anew. If the configuration is still not found in the database, a saddle-point search is carried out to find all the possible processes. This new configuration is then appended to the existing database and the corresponding rate tables in the simulation are updated. In this way initially an empty database is filled up with configurations as they appear during the course of the simulation. During a saddle-point search, part of the system that is larger than the range of interaction is incorporated into the molecular-static calculations for find the activation barriers. This ensures that all the neighboring atoms that could affect the motion of the central atom are included in the search. We note that during saddle point search real coordinates are used.
Since a process involves motion of an atom or atoms from one box to another, to find all the possible processes for a configuration, central atom is dragged in the direction of empty boxes. By dragging the central atom towards the empty box we find all the possible processes that can be obtained using the drag method. In the future we will incorporate other saddle-point searches. 
If the configuration of a central atom is unknown then there is always a possibility that configurations of its neighboring atoms are also unknown, resulting in saddle-point searches for more than one atom. The size of the database usually depends on the system, the range of interaction (or size of the super box), and resolution (or size of the small box) with which a processes need to be identified. The accumulation of this database does not proceed uniformly with time,\cite{slkmc1} but eventually saturates, after which simulation proceeds as a regular KMC with a closed database.

\section{Results}\label{results}
As mentioned earlier that the idea of developing this new pattern recognition is to be able to do simulations where atoms can go off-lattice (non-symmetric sites). As a test application we study homoepitaxial systems, in which atoms move for one on-lattice site to another. 
In this article we examine the decay of 3D Cu islands  on Cu (100) surface and compare results to the existing results in the literature. Also, in order to show that the 3D pattern-recognition scheme is applicable to all types of lattices,  we apply it simulate 2D diffusion of Cu monomer and dimer on Cu(111) surface.  

\subsection{Decay of Cu 3D islands on Cu(100) surface}

\begin{figure}
\center
\subfigure{ \includegraphics [width=6.5cm]{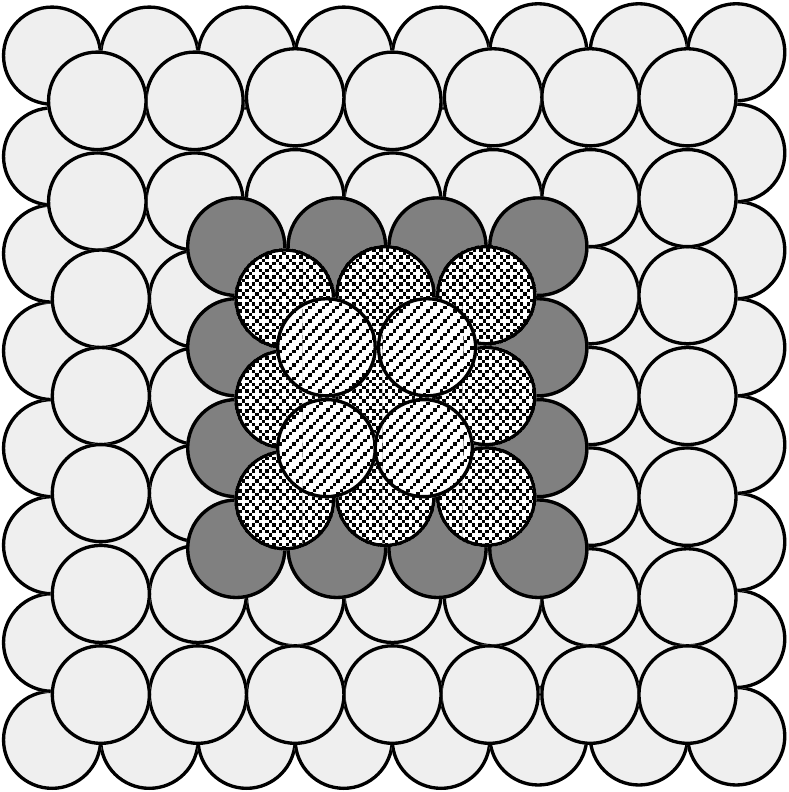}}
\caption{\label{cluster}{Nano-cluster of $29$ atoms}}
\end{figure}

In these simulations we used a fcc(100) substrate with periodic boundary conditions in the x-y plane (parallel to the surface).
3D islands of various sizes, which have a four-sided pyramidal shape with (111) facets (Fig.~\ref{cluster} shows a 3D island of size $29$ atoms) are created manually on this substrate. The decay of these islands is simulated until the entire island reduces to a monolayer and this decay is  studied as a function of temperature and 3D island size. The size of the substrate is $125.44 \times 125.44$ \AA~or $50 \times 50$ lattice units. 

Since our objective here was to test the new pattern-recognition scheme, to save time we have identified the processes and their barriers on the basis of parameterization of EMT barriers \cite{Jacobsen}. We note that this model has previously been used to study multilayer Cu/Cu(100) growth.\cite{Cu_EMT1, Cu_EMT2, FPT}
To take into account the Ehrlich-Schwoebel (ES) barrier to interlayer diffusion,\cite{ES1,ES2} for all interlayer diffusion processes an additional barrier of $0.02$ eV is added to the value of each intra-layer diffusion barrier. 
For atoms at non four-fold hollow sites, our simulation also takes into account downward funneling (DF),\cite{evans1}. Since at the temperatures we examined it can be considered as a non-activated process, we allowed atoms at these sites to ``cascade'' randomly until they find their way to a four-fold hollow site. 
The size of the accumulated database is around $650$ configurations, the actual number depends on the size of the cluster.

\begin{figure}
\center
\subfigure{ \includegraphics [width=6.1cm]{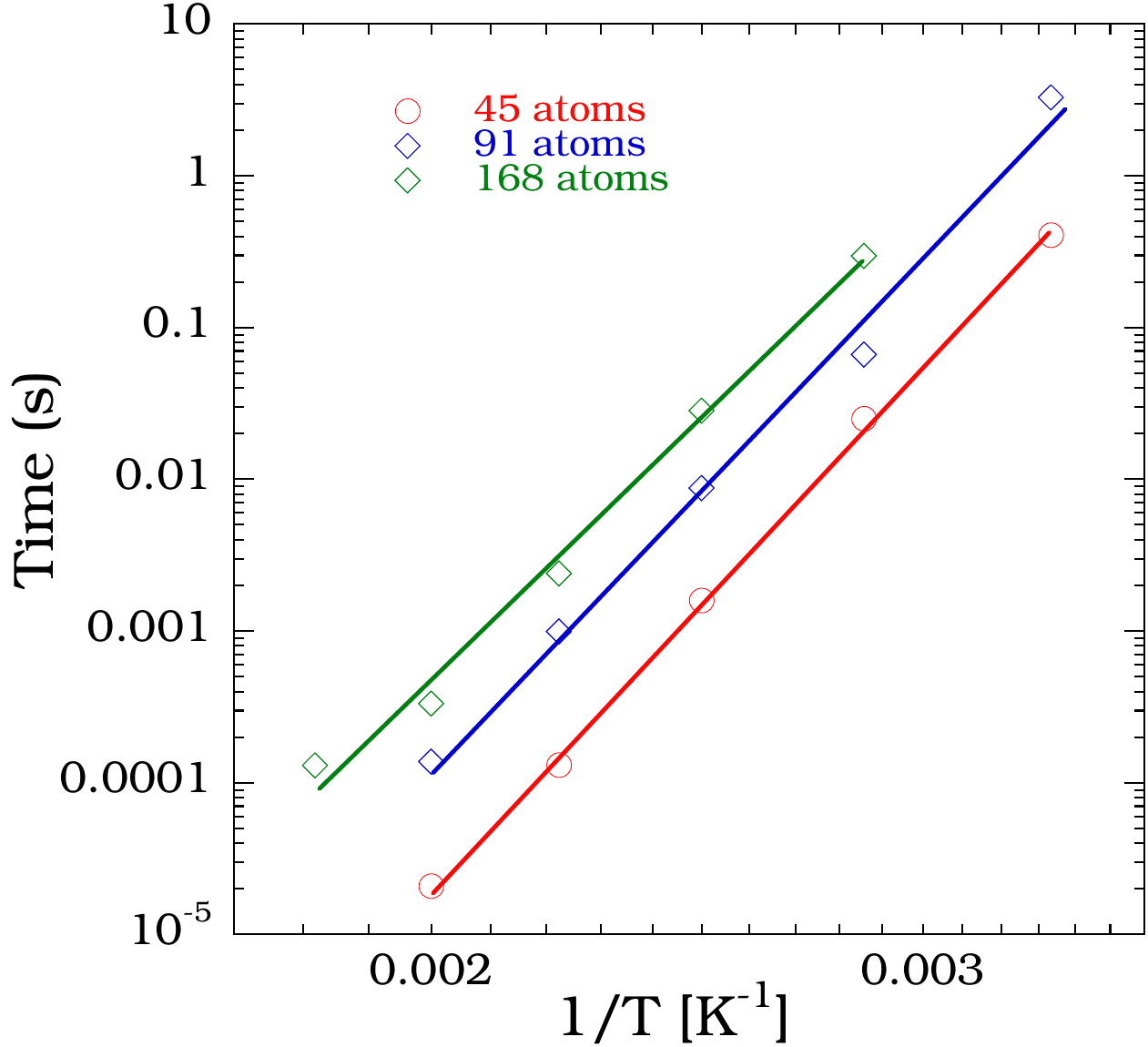}}
\caption{\label{arrh}{The time required for the reduction of 3D nano-clusters to one monolayer for clusters of sizes $45$, $91$ and $68$ atoms at different temperatures.}}
\end{figure}

\begin{figure}
\center
\subfigure{ \includegraphics [width=6.0cm]{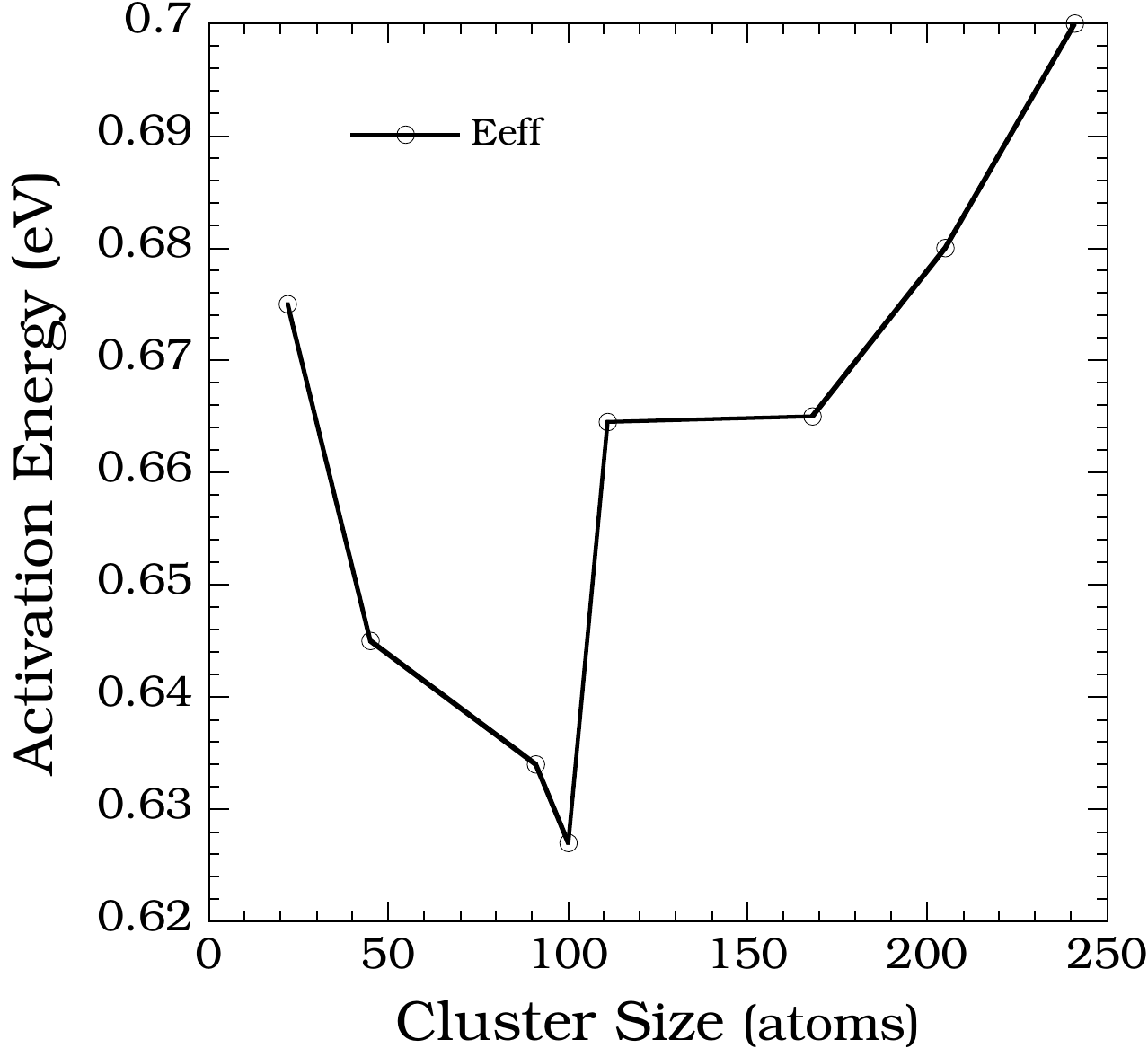}}
\caption{\label{E-eff}{The effective activation energy as a function of the cluster size (number of atoms). This is an effective activation barrier for a 3D-cluster to decay to a monolayer of atoms. Obtained by fitting the data assuming an Arrhenius behavior.}}
\end{figure}

Figure~\ref{arrh} shows the time that is required for 3D islands of sizes $45$, $91$ and $168$ atoms to decay to one monolayer at various temperatures. Effective activation barriers can be calculated by form fitting the data to a straight line assuming an Arrhenius behavior. Fig.~\ref{E-eff} shows the plot of the effective activation barrier as a function of 3D-cluster size with a minimum at  $100$ atom cluster, in agreement with previous published results.\cite{3dcluster}

\subsection{2D diffusion on Cu(111) surface}
The 3D pattern-recognition scheme uses 3D rectangular boxes commensurate with the fcc(100) lattice, which has a square symmetry. To show that this pattern-recognition scheme can be applied to any type of lattice we have also studied 2D diffusion on fcc(111) surface, which has a triangular symmetry. The database of processes for Cu(111) island diffusion was obtained by using the drag method, and activation barriers were checked against those calculated using the more sophisticated nudged-elastic band (NEB) method. The interatomic potentials were modeled using the embedded-atom method.\cite{Foiles} For the sake of simplification we assumed a `normal' value for all diffusion prefactors. Although we are aware that multi-atom processes may be characterized by high prefactors.\cite{handan1,handan2, pref1, pref2}

\begin{figure}
\center
\subfigure{ \includegraphics [width=4.5cm]{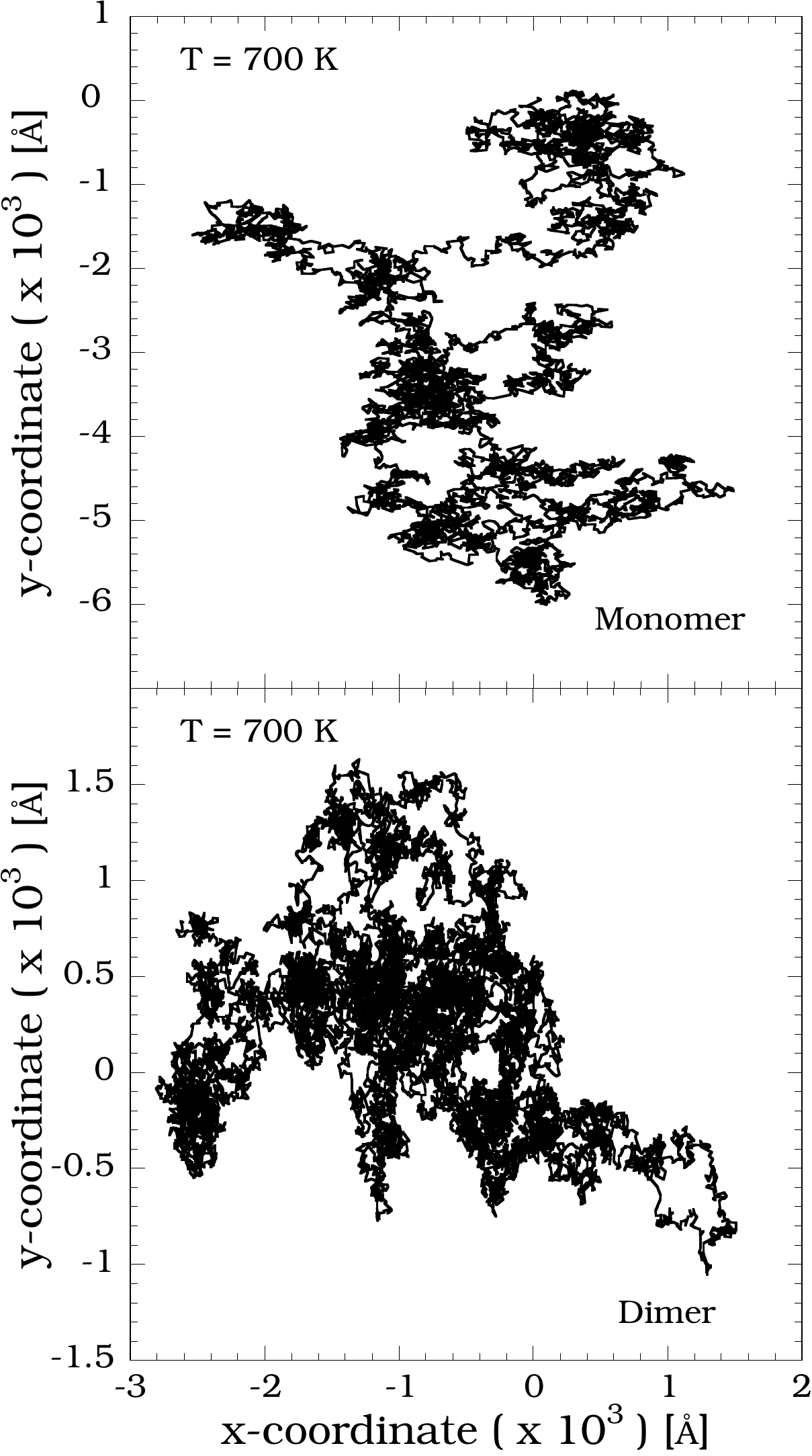}}
\caption{\label{trace}{Trace of the center of mass of Cu monomer and dimer on Cu(111) at $700$ K}}
\end{figure}

In these simulations we place the adatom island on a fcc(111) substrate with periodic boundary conditions in the x-y plane (parallel to the surface) For the fcc (111) surface four layers are used for pattern recognition. These simulations were performed for about $10^{7}$ KMC steps at $300$, $500$ and $700$ K. During each simulation, the position of the center of mass was recorded after every $1000$ KMC steps. The diffusion coefficient $D$ of an adatom island for a 2D random walk was calculated using Einstein Equation:\cite{MSD} $D = \lim_{t \to \infty} \langle R_{CM}(t) - R_{CM}(0)]^{2}\rangle/2dt$, where $R_{CM}(t)$ is the position of the center of mass of the island at time $t$, and  $d$ is the dimensionality of the system. Effective diffusion barriers were extracted for monomer and dimer from Arrhenius plot of $ln~D$  Vs $1/k_{B}T$. The goal was to test the new pattern-recognition scheme by comparing calculated diffusion coefficients and effective energy barriers with already published values.\cite{slkmc2}

\begin{figure}
\center
\subfigure{ \includegraphics [width=4.5cm]{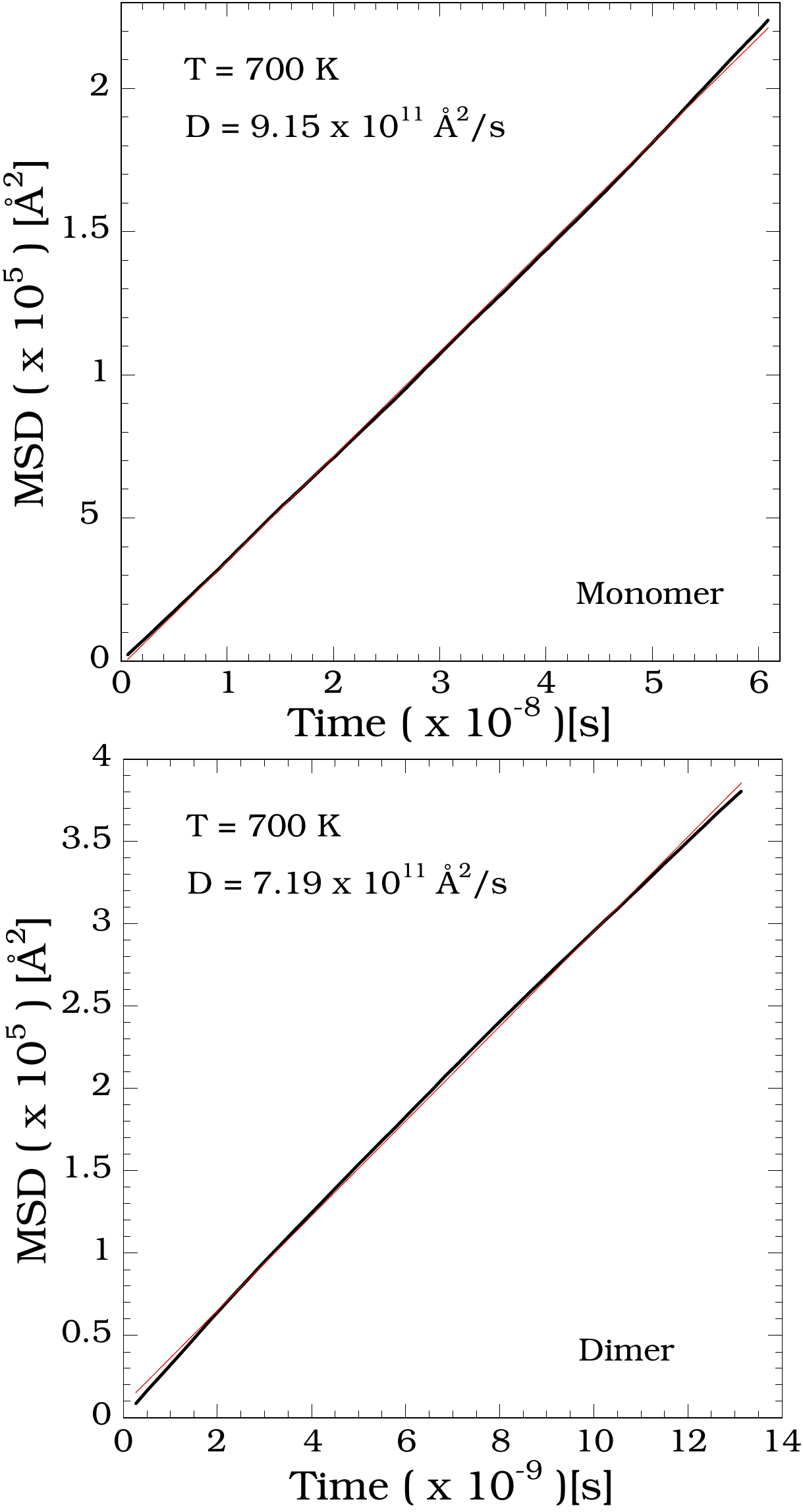}}
\caption{\label{MSD}{Mean square displacement (MSD) for Cu monomer and dimer on Cu(111) as a function of time at $700$ K}}
\end{figure}

 \begin{table}
\caption{\label{table1} Diffusion coefficient for Cu monomer and dimer on Cu(111) ($\rm{\AA^{2}}$/s)}
 \begin{tabular}{c c c c c c}
\hline
\hline
Cluster Size &$300$ K &$500$ K &$700$ K \\
\hline
Monomer &$4.8 \times10^{11}$&$8.2 \times 10^{11}$ &$9.2 \times 10^{11}$\\ 
Dimer &$1.1 \times 10^{11}$&$4.2 \times 10^{11}$&$7.2 \times 10^{11}$\\
 \hline
 \hline
\end{tabular}
\end{table}

In Fig.~\ref{trace}, we show the trace of the position of the center of mass on the x-y plane for both monomer and dimer at $700$ K after every $1000$ KMC steps.  For both monomer and dimer, mean square displacements as a function of time exhibit a linear behavior (within statistical error) and are shown in fig.~\ref{MSD}. The slope extracted from the mean square displacement plot divided by $4$ gives the diffusion coefficient. Table~\ref{table1} shows the diffusion coefficients of monomer and dimer at $4$ different temperatures. The effective barriers extracted from the Arrhenius  plot for the monomer and dimer are $0.029$eV and $0.083$eV, respectively. That these values are within the statistical error of already published results \cite{slkmc2} shows that pattern-recognition scheme works well with 2D diffusion. 

\section{Discussion}\label{conclusions}
In a KMC simulations, processes that can occur are identified based on the local neighborhood of an atom. In traditional KMC simulation this information is hardwired into it and cannot be modified during the course of the simulation. Therefore all the relevant processes that can happen in the system have to be known ahead of time. In SLKMC simulations, generation of unique configuration key to identify a local neighborhood using a pattern-recognition scheme allows storage and retrieval of information about processes from a database. This gives these simulation flexibility to add processes if necessary during the course of the simulation, removing the constraint that all the possible processes have to be known ahead of time.
Accordingly, in order to improve upon previously-developed pattern recognition schemes, which are essentially two-dimensional, and to accurately account for different types of diffusion processes involving 2D or 3D motion of atoms to either symmetric or non-symmetric (off-lattice) sites, we have developed a new 3D off-lattice pattern recognition scheme.  
This new scheme is simple, flexible and more accurate in identifying local neighborhoods. The improvement in accuracy of the scheme results from the matching integers instead of real numbers for pattern recognition while the flexibility results from the application of simulation code without significant changes, to both 2D and 3D systems regardless of whether these are on- or off-lattice. 

\begin{figure}
\center
\subfigure{ \includegraphics [width=5.5cm]{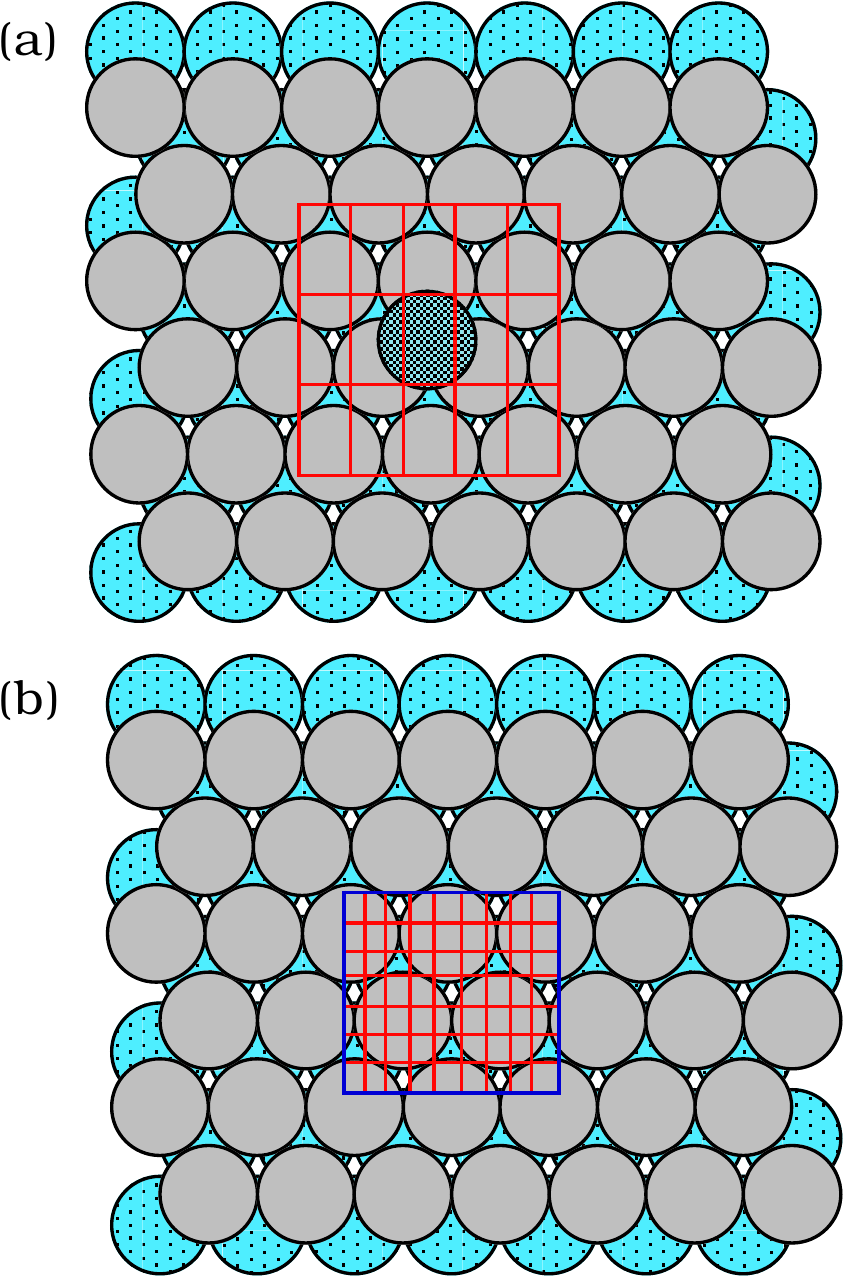}}
\caption{\label{111}{(a) Monomer (dark blue atom) on a fcc site on fcc (111) surface, the red lines show division of space on the third layer $L3$, (b) The red lines show even finer division of space, enabling recognition of configurations involving off-lattice sites}}
\end{figure}

We have tested this method by studying 2D diffusivity of Cu islands on Cu (111) surface and decay of Cu 3D islands on Cu(100) surface and comparing the results with those are available in the literature. 
For Cu/Cu(100) simulations, even though the processes allowed in these simulations move atoms from one symmetric site to another, pattern recognition does allow the identification of neighborhood and corresponding processes if atom is at a non-symmetric (off-lattice) site.  It can be seen from the fig.~\ref{boxfcc_bits} that the division of 3D space allows the identification of a 4-fold hollow site, an A-top site and a bridge site on layer $L2$. 
We note that for fcc (111) surface, two substrate layers and two adatom layers are used for recognizing a pattern.
Fig.~\ref{111}(a) shows division of space for the third layer ($L3$) or adatom layer on fcc (111) surface. It can be seen that the size of the smaller box used is accurate enough to differentiate between fcc and hcp sites. 
This is because in our simulations on fcc(111) surface we allowed atoms to move only from one symmetric site to another. But, the size of the small box  can be easily reduced, thereby increasing number of small boxes, as shown in fig.~\ref{111}(b) to enable the identification of  configurations in which atoms sit on off-lattice sites.
That is, the accuracy with which a process is identified depends on the size of the small box.  Hence some knowledge about the system under study is needed to determine not only the size of the super box but also the dimensions of the small boxes. 

Although this 3D pattern-recognition scheme was developed with the intention of using it to study morphological evolution during heteroepitaxial growth using KMC method, it should be applicable to any 3D system wherever the transition-state theory is applicable.
This new 3D off-lattice pattern recognition scheme is computationally more expensive than previous methods but gives greater flexibility and extends the range of types of systems that can be simulated using KMC method. 
For a fixed size of a super box the computational effort in identifying a pattern does not vary much with number of the small boxes or their size, it increases rather with the increasing size of the super box, because large part of  the computational effort goes into searching the neighborhood atoms the number of which depends on the size of the super box.
We note for the case of 2D island diffusion on fcc (111) surface that each KMC step requires about couple of micro-seconds depending on the computing speed of the processor (On a Intel Core 2 Duo $2.26$ GHz machine, each KMC step takes about $2.2$ms). Thus, these kinds of simulations are an excellent candidate for parallel simulations.


\section{Acknowledgement}
This work was supported by DOE-BES under grant DE-FG02-07ER46354. We would also like to acknowledge of the support of computational resources of University of Central Florida (STOKES) and also grant of computer time from TeraGrid (TG-DMR110046). We thank Lyman Baker for critical reading of the manuscript.

\bibliography{references}

\begin{thebibliography}{10}
\expandafter\ifx\csname url\endcsname\relax
  \def\url#1{\texttt{#1}}\fi
\expandafter\ifx\csname urlprefix\endcsname\relax\def\urlprefix{URL }\fi
\expandafter\ifx\csname href\endcsname\relax
  \def\href#1#2{#2} \def\path#1{#1}\fi

\bibitem{r1}
Y.~Saito, Statistical Physics of Crystal Growth, World Scientific, Singapore,
  1996.

\bibitem{r2}
R.~N{$\ddot {\rm{o}}$}tzel, J.~Temmyo, T.~Tamamura, Nature 369.

\bibitem{r3}
R.~Jullien, J.~Kertesz, P.~Meakin, D.~E. Wolf, Surface Disordering: Growth,
  Roughening and Phase Transitions, Nova, Commack, New York, 1993.

\bibitem{Notomi}
M.~Notomi, J.~Hammersberg, H.~Weman, S.~Nojima, H.~Sugiura, M.~Okamoto,
  T.~Tamamura, M.~Potemski, Phys. Rev. B 52 (1995) 11147.

\bibitem{Floro}
J.~L. Gray, N.~Singh, D.~M. Elzey, R.~Hull, J.~A. Floro, Phys. Rev. Lett. 92
  (2004) 135504.

\bibitem{kordos}
P.~Kordos, J.~Novak (Eds.), Heterostructure Epitaxy and Devices, Boston:
  Kluwer, 1998.

\bibitem{Pearsall}
T.~P.~. Pearsall (Ed.), Strain Layer Super-lattices, Boston: Academic, 1990.

\bibitem{Mo}
Y.~W. Mo, D.~E. Savage, B.~S. Swartzentruber, M.~G. Lagally, Phys. Rev. Lett.
  65 (1990) 1020.

\bibitem{Eaglesham}
D.~J. Eaglesham, M.~Cerullo, Phys. Rev. Lett 64 (1990) 1943.

\bibitem{MD1}
M.~Schneider, A.~Rahman, I.~K. Schuller, Phys. Rev. Lett. 55 (1985) 604.

\bibitem{MD2}
B.~W. Dodson, CRC Crit. Rev. Solid. State Mater. Sci 16 (1990) 115.

\bibitem{MD3}
M.~Schneider, I.~Schuller, A.~Rahman, Phys. Rev. B 36 (1987) 1340.

\bibitem{bkl}
A.~B. Bortz, M.~H. Kalos, J.~L. Lebowitz, J. Comput. Phys. 17 (1975) 10.

\bibitem{gilmer}
G.~H. Gilmer, J. Crystal. Growth. 35 (1976) 15.

\bibitem{voter1}
A.~F. Voter, Phys. Rev. B 34 (1986) 6819.

\bibitem{maksym}
P.~A. Maksym, Semiconf. Sci. Technol. 3 (1988) 594.

\bibitem{kristen}
K.~A. Fichthorn, W.~H. Weinberg, J. Chem. Phys. 95 (1991) 1090.

\bibitem{Blue}
J.~L. Blue, I.~Beichl, F.~Sullivan, Phys. Rev. E 51 (1995) R867.

\bibitem{neb1}
G.~M. H.~J\'{o}nsson, K.~W. Jacobsen, Classical and Quantum Dynamics in
  Condensed Phase Simulations, World Scientific, Singapore, 1998.

\bibitem{henkelman}
G.~Henkelman, H.~Jonsson, J. Chem. Phys. 115 (2001) 9657.

\bibitem{slkmc1}
O.~Trushin, A.~Karim, A.~Kara, T.~S. Rahman, Phys. Rev. B 72 (2005) 115401.

\bibitem{slkmc2}
A.~Karim, A.~N. Al-Rawi, A.~Kara, T.~S. Rahman, O.~Trushin, T.~Ala-Nissila,
  Phys. Rev. B 73 (2006) 165411.

\bibitem{pslkmc}
G.~Nandipati, Y.~Shim, J.~G. Amar, A.~Karim, A.~Kara, T.~S. Rahman, O.~Trushin,
  J. Phys.: Condens. Matter 21~(084214).

\bibitem{iselect}
G.~Nandipati, A.~Kara, S.~I. Shah, T.~S. Rahman, J. Phys.: Condens. Matter 23
  (2011) 262001.

\bibitem{offkmc}
A.~Kara, O.~Trushin, H.~Yildirim, T.~S. Rahman, J. Phys.: Condens. Matter 21.

\bibitem{neb2}
G.~Henkelman, H.~J\'{o}nsson, J. Chem. Phys. 115~(7010).

\bibitem{MD-book}
D.~C. Rapaport, The Art of Molecular Dynamics SImulation, Cambridge University
  Press, 2004.

\bibitem{Jacobsen}
J.~Jacobsen, K.~W. Jacobsen, J.~K. Norskov, Surf. Sci. 359 (1996) 37.

\bibitem{Cu_EMT1}
Y.~Shim, J.~G. Amar, Phys. Rev. B 73.

\bibitem{Cu_EMT2}
J.~Yu, J.~G. Amar, Phys. Rev. Lett 89.

\bibitem{FPT}
G.~Nandipati, Y.~Shim, J.~G. Amar, Phys. Rev. B 81.

\bibitem{ES1}
G.~Ehrlich, F.~G. Hudda, J. Chem. Phys. 44~(1039).

\bibitem{ES2}
L.~Schwoebel, J. Appl. Phys. 40~(614).

\bibitem{evans1}
J.~W. Evans, D.~E. Sanders, P.~A. Thiel, A.~E. DePristo, Phys. Rev. B 41 (1990)
  R5410.

\bibitem{3dcluster}
J.~Frantz, M.~Rusanen, K.~Nordlund, I.~Koponen, J. Phys.: Condens. Matter 16
  (2004) 2995.

\bibitem{Foiles}
S.~M. Foiles, M.~I. Baskes, M.~S. Daw, Phys. Rev. B 33 (1986) 7983.

\bibitem{handan1}
H.~Yildirim, A.~Kara, S.~Durukanoglu, T.~S. Rahman, Surf. Sci. 600 (2006) 484.

\bibitem{handan2}
H.~Yildirim, A.~Kara, T.~S. Rahman, Phys. Rev. B 76 (2007) 165421.

\bibitem{pref1}
G.~Henkelman, H.~J\'{o}nsson, Phys. Rev. Lett. 90 (2003) 116101.

\bibitem{pref2}
F.~Montalenti, Transition-path spectra at metal surfaces, Surf. Sci. 543 (2003)
  141.

\bibitem{MSD}
A.~Einstein, Ann. Phys. (Leipzig) 17 (1905) 549 (English transl. Investigations
  on the Theory of Brownian Movement, Dover, New York, 1956).

\end{thebibliography}

 \end{document}